%% file: ms.tex
\documentclass[runningheads]{llncs}
\usepackage{xcolor}

\usepackage{graphicx}
\usepackage{amsmath}
\usepackage{amssymb}
\usepackage{booktabs}
\usepackage{adjustbox}

\begin{document}

\title{Unsupervised Domain Adaptation Using Feature Disentanglement And GCNs For Medical Image Classification}


\author{Dwarikanath Mahapatra }

\institute{Inception Institute of AI, UAE }

\maketitle

\begin{abstract}
The success of deep learning has set new benchmarks for many medical image analysis tasks. However, deep models often fail to generalize in the presence of distribution shifts between training (source) data and test (target) data. One method commonly employed to counter distribution shifts is domain adaptation: using samples from the target domain to learn to account for shifted distributions. In this work we propose an unsupervised domain adaptation approach that uses graph neural networks and, disentangled semantic and domain invariant structural features, allowing for better performance across distribution shifts. We propose an extension to swapped autoencoders to obtain more discriminative features. We test the proposed method for classification on two challenging medical image datasets with distribution shifts - multi center chest Xray images and histopathology images. Experiments show our method achieves state-of-the-art results compared to other domain adaptation methods. 
  

\end{abstract}

\input{UDA_Intro}

\input{UDA_Method2}

\input{UDA_Expts}

\input{UDA_Concl}


\bibliographystyle{splncs04}

\bibliography{MIDL_UDA_Ref}

\end{document}

%% file: UDA_Intro.tex
\section{Introduction}


The success of convolutional neural networks (CNNs) has set new benchmarks for many medical image classification tasks such as diabetic retinopathy grading \cite{googledr}, digital pathology image classification \cite{googleDP,MahapatraGZSLTMI,LieTMI_2022,Devika_IEEE,MonusacTMI,Mahapatra_Thesis,KuanarVC,MahapatraTMI2021,JuJbhi2020,Frontiers2020,Mahapatra_PR2020} and chest X-ray images \cite{chexpert,NIHXray,MahapatraTIP_RF2014,MahapatraTBME_Pro2014,MahapatraTMI_CD2013,MahapatraJDICD2013,MahapatraJDIMutCont2013,MahapatraJDIGCSP2013,MahapatraJDIJSGR2013,MahapatraTrack_Book,MahapatraJDISkull2012,MahapatraTIP2012,MahapatraTBME2011,MahapatraEURASIP2010,MahapatraTh2012,MahapatraRegBook}, as well as segmentation tasks \cite{Chanti16,Chanti19,ZGe_MTA2019,Behzad_PR2020,Mahapatra_CVIU2019,Mahapatra_CMIG2019,Mahapatra_LME_PR2017,Zilly_CMIG_2016,Mahapatra_SSLAL_CD_CMPB,Mahapatra_SSLAL_Pro_JMI,Mahapatra_LME_CVIU,LiTMI_2015,MahapatraJDI_Cardiac_FSL,Mahapatra_JSTSP2014}. Clinical adoption of algorithms has many challenges due to the domain shift problem: where the target dataset has different characteristics than the source dataset on which the model was trained. These differences are often a result of different image capturing protocols, parameters, devices, scanner manufacturers, etc. 
Since annotating samples from hospitals and domains is challenging due to scarcity of experts and the resource intensive nature of labeling, it is essential to design models from the beginning that perform consistently on images acquired from multiple domains. 

Most approaches to the domain shift problem can be categorized based on the amount of data from the target domain. Fully-supervised domain adaptation (DA) and semi-supervised DA assume availability of a large amount and a small amount of fully-labeled instances from the target domain, respectively, along with data from a source domain \cite{Chanti20,Souryaisbi22,mahapatra2022_midl,Mahapatra_CVAMD2021,PandeyiMIMIC2021,SrivastavaFAIR2021,Mahapatra_DART21b,Mahapatra_DART21a,LieMiccai21,TongDART20,Mahapatra_MICCAI20,Behzad_MICCAI20,Mahapatra_CVPR2020,Kuanar_ICIP19,Bozorgtabar_ICCV19,Xing_MICCAI19,Mahapatra_ISBI19,MahapatraAL_MICCAI18,Mahapatra_MLMI18}. Unsupervised domain adaptation (UDA) techniques \cite{Chanti4,Chanti18,Sedai_OMIA18,Sedai_MLMI18,MahapatraGAN_ISBI18,Sedai_MICCAI17,Mahapatra_MICCAI17,Roy_ISBI17,Roy_DICTA16,Tennakoon_OMIA16,Sedai_OMIA16,Mahapatra_OMIA16,Mahapatra_MLMI16,Sedai_EMBC16,Mahapatra_EMBC16,Mahapatra_MLMI15_Optic,Mahapatra_MLMI15_Prostate,Mahapatra_OMIA15} use only unlabeled data from the target domain and labeled source domain data. UDA methods aim to learn a domain-invariant representation by enforcing some constraint (e.g. Maximum Mean Discrepancy \cite{Chanti15}) that brings the latent space distributions, $z$, of distinct domains closer; ideally leading to more comparable performance in classification/segmentation.

Convolutional neural network (CNN) based state-of-the-art (SOTA) UDA methods often obtain impressive results but typically only enforce alignment of global domain statistics \cite{GCNcvpr_77,MahapatraISBI15_Optic,MahapatraISBI15_JSGR,MahapatraISBI15_CD,KuangAMM14,Mahapatra_ABD2014,Schuffler_ABD2014,Schuffler_ABD2014_2,MahapatraISBI_CD2014,MahapatraMICCAI_CD2013,Schuffler_ABD2013,MahapatraProISBI13,MahapatraRVISBI13,MahapatraWssISBI13,MahapatraCDFssISBI13,MahapatraCDSPIE13,MahapatraABD12,MahapatraMLMI12} resulting in loss of semantic class label information. Semantic transfer methods \cite{GCNcvpr_55,GCNcvpr_58,MahapatraSTACOM12,VosEMBC,MahapatraGRSPIE12,MahapatraMiccaiIAHBD11,MahapatraMiccai11,MahapatraMiccai10,MahapatraICIP10,MahapatraICDIP10a,MahapatraICDIP10b,MahapatraMiccai08,MahapatraISBI08,MahapatraICME08,MahapatraICBME08_Retrieve,MahapatraICBME08_Sal,MahapatraSPIE08,MahapatraICIT06} address this by propagating class label information into deep adversarial adaptation networks. Unfortunately, it is difficult to model and integrate semantic label transfer into existing deep networks.
To deal with the above limitations \cite{MaGCN,IccvGZSl_Ar,ISR_MIDL_Ar,GCN_MIDL_Ar,DevikaAccess_Ar,SouryaISBI_Ar,Covi19_Ar,DARTGZSL_Ar,DARTSyn_Ar,Kuanar_AR2,TMI2021_Ar,Kuanar_AR1,Lie_AR2,Lie_AR,Salad_AR,Stain_AR,DART2020_Ar,CVPR2020_Ar,sZoom_Ar,CVIU_Ar,AMD_OCT,GANReg2_Ar,GANReg1_Ar,PGAN_Ar,Haze_Ar,Xr_Ar,RegGan_Ar,ISR_Ar,LME_Ar,Misc,Health_p,Pat2,Pat3} propose a Graph Convolutional Adversarial Network (GCAN) for unsupervised domain adaptation.
Graph based methods better exploit global relationship between different nodes (or samples) than CNNs and learn both global and local information beneficial for DA.

One way to address domain shift is with feature disentanglement \cite{liu2018unified} by separating latent feature representations into domain invariant (often called structure or content) and domain variant (texture or style) components, thus minimizing the impact of domain variation.
In this work we combine feature disentanglement with graph convolutional networks (GCN) for unsupervised domain adaptation and apply it to two different standard medical imaging datasets for classification and compare it to SOTA methods.

\textbf{Related Work:}
Prior works in UDA focused on medical image classification \cite{AhnTMI20}, object localisation, and lesion segmentation \cite{Ahn_23,Ahn_24}, and histopathology stain normalization \cite{ChangMiccai21}. Heimann et al. \cite{Ahn_23} used GANs to increase the size of training data and demonstrated improved localisation in X-ray fluoroscopy images. Likewise, Kamnitsas et al. \cite{Ahn_24} used GANs for improved lesion segmentation in magnetic resonance imaging (MRI). %
Ahn et al. \cite{AhnTMI20} use a hierarchical unsupervised feature extractor to reduce reliance on annotated training data. Chang et al. \cite{ChangMiccai21} propose a novel stain mix-up for histopathology stain normalization and subsequent UDA for classification.
  Graph networks for UDA \cite{MaGCN,wu2020unsupervised,Pat4,Pat5,Pat6,Pat7,Pat8,Pat9,Pat10,Pat11,Pat12,Pat13,Pat14,Pat15,Pat16,Pat17,Pat18} have been used in medical imaging applications \cite{GraphReview} such as brain surface segmentation \cite{Rev28} and brain image classification \cite{Rev12,Rev175}. However, none of them combine feature disentanglement with GCNs for UDA.

  \textbf{Our Contributions:}
 Although previous works  used feature disentanglement and graph based domain adaptation separately, they did not combine them for medical image classification. 
 %
 %
 %
 In this work: 1) we propose a feature disentanglement module trained to align structural components coupled with a Graph Convolutional Adversarial Network (GCAN) for unsupervised domain adaptation in medical images. 2) We perform feature disentanglement using swapped autoencoders to obtain texture and structural features, which are used for graph construction and defining generator losses. Since, in medical imaging applications, images from the same modality (showing the same organ) have similar texture, we introduce a novel cosine similarity loss components into the loss function that enforces the preservation of the structure component in the latent representation; 3) We demonstrate our method's effectiveness on multiple medical image datasets.
 


%% file: UDA_Method2.tex
\section{Method}
\label{sec:met}

Given a source data set, $\mathcal{D}_S=\{(x^S_i,y^S_i)\}_{i=1}^{n_s}$, consisting of $n_s$ labeled samples, and a target data set, $\mathcal{D}_T=\{(x^T_i)\}_{i=1}^{n_t}$ consisting of $n_t$ unlabeled target samples, unsupervised domain adaptation aims to learn a classifier that can reliably classify unseen target samples. Here, $x^S_i \sim p_S$ is a source data point sampled from source distribution $p_S$, $y^S_i \in \mathcal{Y}_S$ is the label, and $x^T_i \sim p_T$ is a target data point sampled from target distribution $p_T$.
As per the covariate shift assumption, we assume that $p_S(y | x) = p_T(y | x) ~\forall x$. Thus the only thing that changes between source and target is the distribution of the  input samples, $x$.

\begin{figure*}[t]
 \centering
\begin{tabular}{cc}
\includegraphics[height=4.3cm, width=5.5cm]{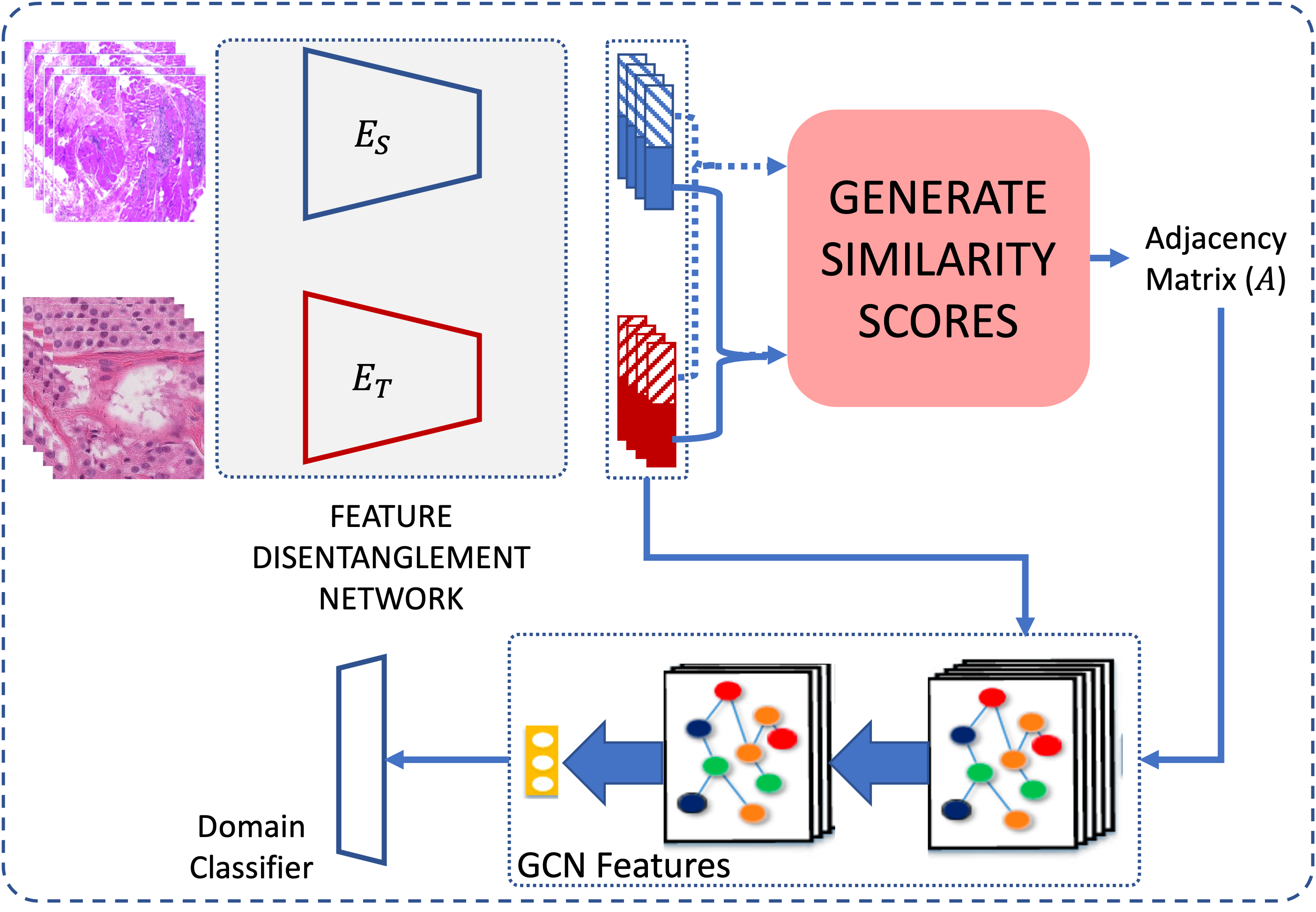} & 
\includegraphics[height=4.3cm, width=6.1cm]{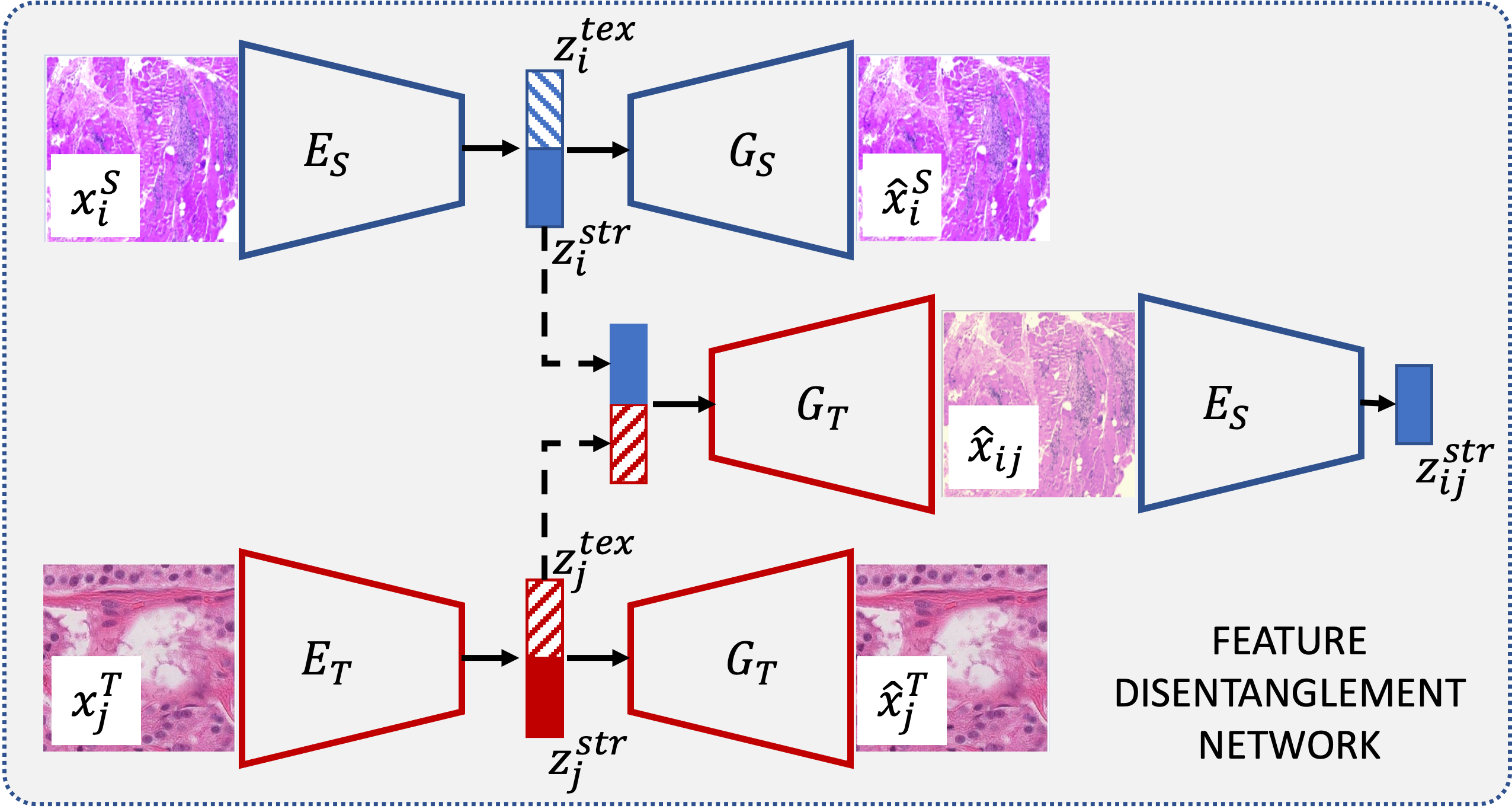} \\
(a) & (b)
\end{tabular}
\caption{(a) Workflow of our proposed method. Given images from source and target domains we disentangle features into texture and structure features, and generate similarity scores to construct an adjacency matrix. Generated features by the GCN are domain invariant and along with the latent features are used to train a domain classifier. (b) Architecture of feature disentanglement network with swapped structure features} 
\label{fig:workflow}
\end{figure*}

The overall block diagram of our system is depicted in Figure \ref{fig:workflow}. Our proposed approach consists of a feature disentanglement module (Figure \ref{fig:workflow}-b) which separates semantic textural and structural features (often called style and content respectively). The output of the feature disentanglement module is constructed into a graph and then fed into an adversarial generator based graph neural network (Figure \ref{fig:workflow}-a) that generates features which are domain invariant. The graph neural network learns more global relationships between samples, which in turn leads to more discriminative and domain invariant feature learning.

\subsection{Feature Disentanglement Network:} 

Figure~\ref{fig:workflow}(b) shows the architecture of our feature disentanglement network (FDN). The FDN consists of two encoders ($E_S(\cdot), E_T(\cdot)$) and two decoder (generator) networks ($G_S(\cdot), G_T(\cdot)$), for the source and target domains respectively.  Similar to a classic autoencoder, each encoder, $E_\bullet(\cdot)$, produces a latent code $z_i$ for image $x_i^\bullet \sim p_\bullet$. Each decoder, $G_\bullet(\cdot)$, reconstructs the original image from $z_i$. Furthermore,  we  divide  the  latent code, $z_i$, into two  components: a texture component, $z^{tex}_i$, and a structural component, $z^{str}_i$. Structural information is the local information that describes an image's underlying structure and visible organs or parts, and is expected to be invariant across similar images from different domains. The disentanglement network is trained using the following loss function:
\begin{equation}
    \mathcal{L}_{Disent} =\mathcal{L}_{Rec} + \lambda_1\mathcal{L}_{swap} + \lambda_2\mathcal{L}_{str} 
    \label{eq:loss2}
\end{equation}
%
$\mathcal{L}_{Rec}$, is the commonly used image reconstruction loss and is defined as:
\begin{equation}
    \mathcal{L}_{Rec}=\mathbb{E}_{x_i^S\sim p_{S}}\left[\left\|x_i^S-G_S(E_S(x_i^S)) \right\| \right] + \mathbb{E}_{x_j^T\sim p_{T}}\left[\left\|x_j^T-G_T(E_T(x_j^T)) \right\| \right]
\end{equation}
As shown in Figure~\ref{fig:workflow}(b), we also combine the structure of source domain images ($z^{str}_i$) with the texture of target images ($z^{tex}_j$) and send it though $G_T$. the reconstructed images, $\hat{x}_{ij}$, should contain the structure of the original source domain image but, should look like an image from the target domain. To enforce this property, we use the adversarial loss term $\mathcal{L}_{swap}$ defined as:
\begin{equation}
    \mathcal{L}_{swap} = \mathbb{E}_{x_j \sim p_T} [\log(1-D_{FD}(x_j))]  + \mathbb{E}_{x_i\sim p^S,x_j \sim p^T} \left[\log(D_{FD}(G_T([z_i^{str},z_j^{tex}]))) \right].
\end{equation}
Here, $D$ is a discriminator trained with target domain images, thus the generator should learn to produce images that appear to be from the target domain. 

Furthermore, the newly generated image, $\hat{x}_{ij}$, should retain the structure of the original source domain image $x_i^S$ from where the original structure was obtained. In  \cite{SwapVAE} the authors use co-occurrent patch statistics from $\hat{x}_{ij}$ and $x_i^S$ such that any patch from $x_i^S$ cannot be distinguished from a group of patches from $\hat{x}_{ij}$. This constraint is feasible in natural images since different images have far more varied style and content, but is challenging to implement in medical images since images from the same modality (showing the same organ) appear similar. Thus limited unique features are learnt under such a constraint. 
To overcome this limitation we introduce a novel component into the loss function. The output, $\hat{x}_{ij}$, when passed through encoder $E_S$ produces the disentangled features $z^{str}_{ij},z^{tex}_{ij}$. Since $\hat{x}_{ij}$ is generated from the combination of $z^{str}_i$ and $z^{tex}_j$, $z^{str}_{ij}$ should be similar to $z^{str}_i$ (i.e. the structure should not change). 
This is enforced using the cosine similarity loss 
\begin{equation}
    \mathcal{L}_{str} = 1-\langle z^{str}_i,z^{str}_{ij}\rangle
    \label{eq:disloss2}
\end{equation}
where $\langle \cdot \rangle$ denotes cosine similarity. 

\subsection{Graph Convolutional Adversarial Network}


To learn more global relationships between samples, the features from the feature disentanglement module is used to construct an instance graph and which is processed with a Graph Convolutional Network (GCN).
%
%
The GCN learns layer-wise propagation operations that can be applied directly on graphs.
Given an undirected graph with $m$ nodes, the set of edges
between nodes, and an adjacency matrix $\textbf{A} \in R^{m\times m}$, a linear transformation  of graph convolution is defined  as the multiplication of a graph signal $\textbf{X} \in R^{k\times m}$ with a filter $\textbf{W} \in R^{k\times c}$ :

\begin{equation}
    \textbf{Z}=\hat{\textbf{D}}^{-\frac{1}{2}} \hat{\textbf{A}} \hat{\textbf{D}}^{-\frac{1}{2}}  \textbf{X}^{T}\textbf{W},
    \label{eq:GCNscore}
\end{equation}
where $\hat{\textbf{A}}=\textbf{A}+\textbf{I}$, $\textbf{I}$ being the identity matrix, and $\hat{\textbf{D}_{ii}}=\sum_j \hat{\textbf{A}}_{ij}$. The output  is a $c\times m$ matrix $\textbf{Z}$. The GCN can be constructed by stacking multiple graph convolutional layers followed by a non-linear operation (such as ReLU).

In our implementation, each node in the instance graph
represents a latent feature of a sample (either source or target domain) $[z^{tex},z^{str}]$. The adjacency matrix is constructed using the cosine similarity score as 
$\hat{\textbf{A}}=\textbf{X}_{sc}\textbf{X}_{sc}^T$.
Note that the cosine similarity score quantifies the semantic similarity between two samples
. Here $\textbf{X}_{sc}\in R^{w\times h}$ is the matrix of cosine similarity scores, $w$ is the batch size, and $h=2$ is the dimension of the similarity scores of each sample obtained from corresponding  $[z_{tex},z_{str}]$.

The flow of our proposed approach is shown in Figure \ref{fig:workflow}(a). 
Our network is trained by minimizing the following objective function:
\begin{equation}
    \mathcal{L}(\mathcal{X}_S,\mathcal{Y}_S,\mathcal{X}_T)=\mathcal{L}_C(\mathcal{X}_S,\mathcal{Y}_S) + \lambda_{adv}\mathcal{L}_{Adv}(\mathcal{X}_S,\mathcal{X}_T) 
\end{equation}
The classification loss $\mathcal{L}_C$ is a cross entropy loss  
%
$\mathcal{L}_C(\mathcal{X}_S,\mathcal{Y}_S)=-\sum_{c=1}^{n_s} y_c \log(p_c)$, 
  $y_c$ is the indicator variable and $p_c$ is the class probability. 


The second loss component is an adversarial  loss function defined in Eq.~\ref{eq:advloss1}. A domain classifier $D$ identifies whether features from the graph neural network $G_{GNN}$ are from the source or target domain. Conversely, $G_{GNN}$, is being trained to produce samples that can fool $D$. This
two-player minimax game will reach an equilibrium state when features from $G_{GNN}$ are domain-invariant.
\begin{equation}
    \mathcal{L}_{Adv}(\mathcal{X}_S,\mathcal{X}_T) =\mathbb{E}_{x\in D_S} [\log(1-D(G_{GNN}(x)))] + \mathbb{E}_{x\in D_T} [\log(D(G_{GNN}(x)))]
    \label{eq:advloss1}
\end{equation}



\textbf{Training And Implementation:}
Given $\textbf{X}$ and $\textbf{A}$ the GCN features are obtained according to Eq.\ref{eq:GCNscore}. Source and target domain graphs are individually constructed and fed into the parameters-shared GCNs to learn representations.  The dimension of $z_{tex}$ is $256$, while $z_{str}$ is $64\times64$. 
\textbf{VAE Network:} The encoder consists of $3$ convolution blocks followed by max pooling. The decoder is symmetrically designed to the encoder. $3\times3$ convolution filters are used and $64,32,32$ filters are used in each conv layer. The input to the VAE is $256\times256$. For the domain classifier we use DenseNet-121 network with pre-trained weights from ImageNet and fine-tuned using self supervised learning. As a pre-text task we use the classifier to predict the intensity values of masked regions.

%% file: UDA_Expts.tex
\section{Experiments And Results}


\subsection{Results For CAMELYON17 Dataset}

\textbf{Dataset Description:}
 We use the CAMELYON17  dataset \cite{CAMELYON17} to evaluate the performance of the proposed method on tumor/normal classification. In this dataset, a
total of 500 $H\&E$ stained WSIs are collected from five medical centers (denoted as by $C1_{17},C2_{17},C3_{17},C4_{17}$, $C5_{17}$ respectively). $50$ of these WSIs include lesion-level annotations. All
positive and negative WSIs are randomly split into training/validation/test sets and provided by the organizers in a $50/30/20 \%$ split for the individual medical centers to obtain the following split: $C1_{17}$:37/22/15, $C2_{17}$: 34/20/14, $C3_{17}$: 43/24/18, $C4_{17}$: 35/20/15, $C5_{17}$: 36/20/15. $256\times256$ image patches are extracted from  the annotated tumors for positive patches and from tissue regions of WSIs without tumors for negative patches. We use $\lambda_1=0.9,\lambda_2=1.2,\lambda_{adv}=1.3$. 

\textbf{Implementation Details}
Since the images have been taken from different medical centers their appearance varies despite sharing the same disease labels. This is due to slightly different protocols of $H\&E$ staining. Stain normalization has been a widely explored topic which aims to standardize the appearance of images across all centers, which is equivalent to domain adaptation. Recent approaches to stain normalization/domain adaptation favour use of GANs and other deep learning methods.  We compare our approach to recent approaches and also with \cite{ChangMiccai21} which explicitly performs UDA using MixUp.

To evaluate our method's performance: 1) We use $C1_{17}$ as the source dataset and train a ResNet-50 classifier \cite{ResNet} (ResNet$_{C1}$). Each remaining  dataset from the other centers are, separately, taken as the target dataset, the corresponding domain adapted images are generated, and classified using ResNet$_{C1}$. As a baseline, we perform the experiment without domain adaptation denoted as $No-UDA$ where ResNet$_{C1}$ is used to classify images from other centers. We report results for a network trained in a  fully-supervised manner on the training set from the same domain ($FSL-Same Domain$) to give an  upper-bound expectation for a UDA model trained on other domain's data, where a ResNet-50 is trained on the training images and used to classify test images, all from the same hospital. This approach will give the best results for a given classifier (in our case ResNet-50). All the above  experiments are repeated using each of $C2_{17},C3_{17},C4_{17},C5_{17}$ as the source dataset.
Table~\ref{tab:cam17} summarizes our results. PyTorch was used in all implementations.

The results in Table~\ref{tab:cam17} show that UDA methods are  better than conventional stain normalization approaches as evidenced by the superior performance of our proposed method and \cite{ChangMiccai21}. Our method performs the best amongst all the methods, and approaches $FSL-Same Domain$, the theoretical maximum performance a domain adaptation method can achieve. This shows that our proposed GCN based approach performs better than other UDA methods. The ablation studies also show that our proposed individual loss term $\mathcal{L}_{Str}$ has a significant contribution to the overall performance of our method and excluding it significantly degrades the performance.

\begin{table}[t]
 \begin{center}
 \begin{adjustbox}{width=\textwidth}
\begin{tabular}{|c|c|c|c|c|c|c|c|c|c|c|c|}
\hline 
{} & \multicolumn{7}{c|}{Comparison and Proposed Method} & {\textbf{Proposed}} & \multicolumn{3}{|c|}{Ablation Study Results} \\ \hline
{} &  {No UDA} & {{MMD}} & {CycleGAN} & {\cite{Vahadane}} & {\cite{gadermayr2018miccai}} & {\cite{MahapatraMiccai20}} & {\cite{ChangMiccai21}} & {\textbf{Method}}   & {FSL-Same} & { $\mathcal{L}_{Str}$} & $\mathcal{L}_{Swap}$ \\ \hline
{$Center~1$} & {0.8068} & {{0.8742}} & {0.9010} & {0.9123} & {0.9487}& {{0.9668}} & {0.979} & {\textbf{0.988}}  & {0.991} & {0.979} & {0.953}  \\ \hline
{$Center~2$}  & {0.7203} & {{0.6926}} & {0.7173} & {0.7347} & {0.8115} & {{0.8537}}  & {0.948}  & {\textbf{0.963}}  & {0.972} & {0.951} & {0.941}  \\ \hline
{$Center~3$} & {0.7027}  & {0.8711}  & {0.8914} & {0.9063} & {0.8727} & {{0.9385}}   & {0.946}  & {\textbf{0.958}} & {0.965} & {0.948} & {0.932}  \\ \hline
{$Center~4$} & {0.8289} & {0.8578} & {0.8811}  & {0.8949} & {0.9235}  & {{0.9548}}  & {0.965}  & {\textbf{0.979}}   & {0.986} & {0.967} & {0.949} \\ \hline
{$Center~5$} & {0.8203} & {0.7854} & {0.8102} & {0.8223} & {0.9351}  & {{0.9462}} & {0.942}  & {\textbf{0.949}} & {0.957} & {0.934} & {0.921}  \\ \hline
{$Average$} & {0.7758} & {0.8162}  & {0.8402}& {0.8541} & {0.8983} & {{0.9320}}  & {0.956}  & {\textbf{0.967}}  & {0.974} &  {0.956} & {0.939}  \\ \hline
{$p$} & {0.0001} & {0.0001} & {0.002} & {0.003} & {0.013} & {0.024}  & {0.017} &  {-} & {0.07}  & {0.01} & {0.001} \\ \hline
%
\end{tabular}
\end{adjustbox}
\caption{Classification results in terms of AUC measures for different domain adaptation methods on the CAMELYON17 dataset. Note: $FSL-SD$ is a fully-supervised model trained on target domain data. }
\label{tab:cam17}
\end{center}
\end{table}


\subsection{Results on Chest Xray Dataset}

%
We use the following chest Xray datasets:
   \textbf{NIH Chest Xray} Dataset: The NIH ChestXray14 dataset \cite{NIHXray} has $112,120$ expert-annotated frontal-view X-rays from $30,805$ unique patients and has $14$ disease labels. Original images were resized to $256\times256$, and $\lambda_1=0.9,\lambda_2=1.2,\lambda_{adv}=1.2$. 
   \textbf{CheXpert} Dataset: This datset \cite{chexpert} has $224,316$ chest radiographs of $65,240$ patients labeled for the presence of $14$ common chest conditions. 
    The validation ground-truth is obtained using majority voting from annotations of $3$ board-certified radiologists. Original images were resized to $256\times256$, and $\lambda_1=0.95,\lambda_2=1.1,\lambda_{adv}=1.3$. These two datasets have the same set of disease labels. 
    

We divide both datasets into train/validation/test splits on the patient level at $70/10/20$ ratio, such that images from one patient are in only one of the splits. Then we train a DenseNet-121 \cite{CheXNet} classifier on one dataset (say NIH's train split). Here the NIH dataset serves as the source data and CheXpert is the target dataset. We then apply the trained model on the training split of the NIH dataset and tested on the test split of the same domain the results are denoted as $FSL-Same$. When we apply this model to the test split of the CheXpert data without domain adaptation the results are reported under No-$UDA$.

Tables~\ref{tab:UDAXray1},\ref{tab:UDAXray2} show classification results for different DA techniques   where, respectively, the NIH and CheXpert dataset were the source domain and the performance metrics are for, respectively, CheXpert and NIH dataset's \emph{test split}. 
We observe that UDA methods perform worse than $FSL-Same$. This is expected since it is very challenging to perfectly account for domain shift. However all UDA methods perform better than fully supervised methods trained on one domain and applied on another without domain adaptation.

The DANN architecture \cite{DANN} outperforms MMD and cycleGANs, and is on par with graph convolutional methods GCAN \cite{MaGCN} and GCN2 \cite{Rev175}. However our method outperforms all compared methods which can be attributed to the combination of GCNs, which learn more useful global relationships between different samples, and feature disentanglement which in turn leads to more discriminative feature learning.

\begin{table}[ht]
\begin{center}
 \begin{adjustbox}{width=\textwidth}
\begin{tabular}{|c|c|c|c|c|c|c|c|c|c|c|c|c|c|c|}
\hline
{} & Atel. & Card. & Eff. & Infil. & Mass & Nodule & Pneu. & Pneumot. & Consol. &  Edema & Emphy. & Fibr. & PT. & Hernia \\ \hline
{No UDA} & {0.697} & {0.814} & {0.761} & {0.652} & {0.739}  & {0.694} & {0.703} &  {0.781} & {0.704}  & {0.792} & {0.815} & {0.719} & {0.728} & {0.811}     \\ \hline
{MMD} & {0.741} & {0.851} & {0.801} & {0.699} & {0.785} & {0.738} & {0.748} & {0.807} & {0.724} & {0.816} & {0.831} & {0.745} & {0.754}  & {0.846} \\ \hline
{CycleGANs} & {0.765} & {0.874} & {0.824} & {0.736} & {0.817} & {0.758} & {0.769} & {0.832} & {0.742} & {0.838} & {0.865} & {0.762} & {0.773} & {0.864}  \\ \hline
{DANN} & {0.792} & {0.902} & {0.851} & {0.761} & {0.849} & {0.791} & {0.802} & {0.869} & {0.783} & {0.862} & {0.894} & {0.797} & {0.804} & {0.892} \\ \hline
{FSL$-Same$} & \textit{0.849} & \textit{0.954} & \textit{0.903} & \textit{0.814} & \textit{0.907} & \textit{0.825} & \textit{0.844} & \textit{0.928} & \textit{0.835} & \textit{0.928} & \textit{0.951} & \textit{0.847} & \textit{0.842} &  \textit{0.941} \\  \hline
{GCAN} & {0.798} & {0.908}  & {0.862} & {0.757} & {0.858} & {0.803} & {0.800} & {0.867} & {0.776}  & {0.865} & {0.908} & {0.811} & {0.799} &  {0.904}   \\ \hline
{GCN2} & {0.809} & {0.919} & {0.870} & {0.765} & {0.871} & {0.807} & {0.810} & {0.882} & {0.792} & {0.883} & {0.921} & {0.817} & {0.812} & {0.914}     \\ \hline
{Proposed} & \textbf{0.825} & \textbf{0.931}  & \textbf{0.884} & \textbf{0.788} & \textbf{0.890} & \textbf{0.818} & \textbf{0.828} & \textbf{0.903} & \textbf{0.818} & \textbf{0.910} & \textbf{0.934} & \textbf{0.828} & \textbf{0.830} &   \textbf{0.923} \\  \hline
\multicolumn{15}{c}{Ablation Study Results} \\ \hline
{$\mathcal{L}_{Str}$} & {0.811} & {0.914} & {0.870} & {0.771} & {0.872} & {0.808} & {0.819} & {0.884} & {0.803} & {0.887} & {0.916} & {0.809} & {0.812} &  {0.907}  \\ \hline
{$\mathcal{L}_{Swap}$}                     & {0.787} & {0.885} & {0.845} & {0.741} & {0.843} & {0.780} & {0.791} & {0.856} & {0.776} & {0.859} & {0.887} & {0.785} & {0.786} &  {0.883}  \\ \hline
%
%
\end{tabular}
\end{adjustbox}
\caption{Classification results on the CheXpert dataset's test split using NIH data as the source domain. Note: $FSL-SD$ is a fully-supervised model trained on target domain data. }
\label{tab:UDAXray1}
\end{center}
\end{table}

\begin{table}[h]
\begin{center}
 \begin{adjustbox}{width=\textwidth}
\begin{tabular}{|c|c|c|c|c|c|c|c|c|c|c|c|c|c|c|}
\hline
 & Atel. & {Card.}  & {Eff.} & {Infil.} & {Mass} & {Nodule} & {Pneu.} & {Pneumot.} & {Consol.} & {Edema} & {Emphy.} & {Fibr.} & {PT} & {Hernia} \\ \hline
{No UDA} & {0.718} & {0.823} & {0.744} & {0.730} & {0.739} & {0.694} & {0.683} & {0.771} & {0.712} & {0.783} & {0.803} & {0.711} & {0.710} & {0.785} \\ \hline
{MMD} & {0.734} & {0.846} & {0.762} & {0.741} & {0.785} & {0.738} & {0.709} & {0.793} & {0.731} & {0.801} & {0.821} & {0.726} & {0.721} & {0.816} \\ \hline
{CycleGANs} & {0.751} & {0.861} & {0.785} & {0.761} & {0.817} & {0.758} & {0.726} & {0.814} & {0.746}  & {0.818} & {0.837} & {0.741} & {0.737} & {0.836}   \\ \hline
{DANN} & 0.773 & 0.882 & 0.819 & 0.785 & 0.837 & 0.791 & 0.759 & 0.838 & 0.770 & 0.836  & 0.863 & 0.766 & 0.762 & 0.861 \\ \hline
{FSL$-Same$} & \textit{0.814} & \textit{0.929} & \textit{0.863} & \textit{0.821} & \textit{0.869} & \textit{0.825} & \textit{0.798} & \textit{0.863} & \textit{0.805} & \textit{0.872} & \textit{0.904} & \textit{0.802} & \textit{0.798} & \textit{0.892} \\ \hline
{GCAN} & {0.78} & 0.895 & 0.811 & 0.777 & 0.828 & 0.782 & 0.751 & 0.832 & 0.765 & 0.828 & 0.857 & 0.761 & 0.756 & 0.851 \\ \hline
GCN2  & 0.786 & {0.906} & 0.833 & 0.789 & 0.831 & 0.802 & 0.763 & 0.835 & 0.774 & 0.837 & 0.868 & 0.768 & 0.763 & 0.860\\ \hline
{Proposed} & \textbf{0.798} & \textbf{0.931} & \textbf{0.884} & \textbf{0.799} & \textbf{0.843} & \textbf{0.818} & \textbf{0.773} & \textbf{0.847} & \textbf{0.784} & \textbf{0.849} & \textbf{0.879} & \textbf{0.779} & \textbf{0.774} & \textbf{0.868} \\ \hline
\multicolumn{15}{c}{Ablation Study Results} \\ \hline
{$\mathcal{L}_{Str}$} & {0.788} & {0.919} & {0.875} & {0.786} & {0.832} & {0.804} & {0.768} & {0.834} & {0.778} & {0.835} & {0.863} & {0.766} & {0.765} &  {0.853}  \\ \hline
{$\mathcal{L}_{Swap}$}                    & {0.770} & {0.903} & {0.860} & {0.771} & {0.811} & {0.788} & {0.748} & {0.814} & {0.761} & {0.816} & {0.845} & {0.746} & {0.749} &  {0.836}  \\ \hline
%
%
\end{tabular}
\end{adjustbox}
\caption{Classification results on the NIH Xray dataset's test split using CheXpert data as the source domain. Note: $FSL-SD$ is a fully-supervised model trained on target domain data. }
\label{tab:UDAXray2}
\end{center}
\end{table}

%% file: UDA_Concl.tex
%
\section{Conclusion}

In this paper we  propose an UDA method using graph adversarial convolutional networks and feature disentanglement using a novel loss term. Our novel loss term extends the swapped autoencoder architecture thus learning more discriminative features. 
%
 Our graph convolutional network based unsupervised domain adaptation method outperforms conventional CNN  methods as graphs better learn the interaction between samples by focusing on more global interactions while CNNs focus on the local neighborhood. This enables GCN to perform better  UDA as  demonstrated by results on multiple datasets. 
 While feature disentanglement also contributes to improved performance, there is scope for improvement. In future work we wish to explore stronger disentanglement techniques, and aim to extend this approach for segmentation-based tasks.


